\documentclass[a4paper]{jpconf}
\usepackage{graphicx}

\usepackage{wrapfig} 

\begin{document}

\title{Dissipative dynamics of superconducting hybrid qubit systems}
\author{Enrique Montes, Jes\'us M Calero and John H Reina}

\address{Departamento de F\'isica, Universidad del Valle, A.A. 25360, Cali, Colombia}
\ead{enriquem@univalle.edu.co and jhreina@univalle.edu.co}

\begin{abstract}
We perform a theoretical study of composite superconducting qubit systems for
the case of a coupled qubit configuration based on a hybrid qubit circuit
made of both charge and phase qubits, which are coupled via a $%
\sigma_x\otimes\sigma_z$ interaction. We compute the system's
eigen-energies in terms of the qubit transition frequencies and the strength of
the inter-qubit coupling, and describe the sensitivity of the energy
crossing/anti-crossing features to such coupling. We compute the
hybrid system's dissipative dynamics for the cases of i) collective and ii)
independent decoherence, whereby the system interacts with one common and
two different baths of harmonic oscillators, respectively. The calculations
have been performed within the Bloch-Redfield formalism and we report the
solutions for the populations and the coherences of the system's reduced
density matrix. The dephasing and relaxation rates are explicitly calculated
as a function of the heat bath temperature. 
\end{abstract}
\vspace{-0.3cm}
\noindent
To appear  in {\jpcs}
\vspace{-0.2cm}
\section{Introduction}

A quantum computer uses quantum two-level systems or quantum bits (qubits) which must be essentially interacting systems weakly or, even more desirable, completely
decoupled from dissipative environments \cite{intro}. The low temperature
superconductors, like aluminum, are good candidates for this purpose. For
this reason, in the last few years, interest in the study of
superconducting-qubit systems has grown \cite{mak, nak}. In these kind of systems, the
circuit must be cooled to a temperature at which the thermal energy $\beta^{-1}\equiv k_BT$ is
less than the transition energy $\hbar {\omega _{0,1}}$ between the qubit basis
states, say  $\left\vert 0\right\rangle$ and $\left\vert 1\right\rangle$. Another important requirement is that $\hbar {\omega _{0,1}}<<\Delta ,$
where $\Delta $ is the energy gap of the superconducting material \cite%
{solid}. The basic elements used in the preparation of superconducting
qubits are the Josephson junctions, which are one of the few electronic
systems which, besides being nonlinear, also exhibit the property of
a very weak dissipation \cite{state}. Quantum information processors require a qubit coherent dynamics in order to allow the construction of quantum logic gates and circuits \cite{intro}. The hybrid systems have had an impact in the development of
these circuits because they can bind the scales of atomic
motion to macroscopic degrees of freedom \cite{neel}. Some solid-state
systems, such as the so-called superconducting qubits, allow the implementation of such hybrid systems in
their respective circuits which can be connected in different ways, in the form of 
simple circuit elements \cite{xiao}.


\section{Model}


{\it Hybrid quantum circuit.---} 
A generic superconducting qubit can be modeled by the Hamiltonian
\begin{equation}
\widehat{H}_{qb}=-\frac{1}{2}\epsilon {\widehat{\sigma }_{z}}-\frac{1}{2}%
\Delta {\widehat{\sigma }_{x}},  \label{HU}
\end{equation}%
\begin{wrapfigure}{r}{0.52\textwidth}
\vspace{-5pt}
\hspace{-8pt}
\includegraphics*[width=0.53\textwidth]{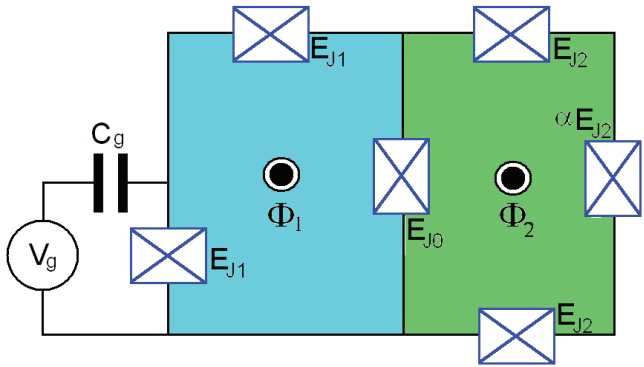}
\vspace{-18pt}
\caption{ Hybrid quantum circuit: charge (left)  and flux (right) qubits   are effectively coupled due to the Josephson junction $E_{J0}$. $E_{Ji}$ denotes the Josephson energy of each junction (crossed boxes). The  charge ($i=1$) and the flux ($i=2$)  qubits are crossed by externally controlled magnetic fluxs $\Phi_{i}$.}
\vspace{-10pt}
\label{circu}
\end{wrapfigure}
where $\epsilon $ and $\Delta $ denote the longitudinal and transversal
parameters of the corresponding qubit, and $\sigma_i$ are the usual Pauli matrices. Charge and flux qubits are two different types of superconducting qubits for
quantum computing. In short, the charge qubit has the advantage of a more flexible
controllability via external parameters: it can be conveniently controlled
by a voltage gate  or an applied magnetic flux. These external control parameters 
appear in the longitudinal ($\sigma _{z}$) and transverse ($%
\sigma _{x}$) terms of the circuit's reduced Hamiltonian. For  the flux
qubit,  the longitudinal term can be controlled by the applied magnetic flux,
but it is hard to control the transversal term via an external parameter. 

The hybrid 
 quantum circuit we study in this work is a system that couples a charge and a flux qubit, as schematically shown in Figure \ref{circu}, 
with an effective interaction due to a Josephson junction that binds them. This
system has been proposed in order to control the transversal term of the flux qubit with
the charge qubit \cite{xiao}. A calculation of the explicit Hamiltonian for  the hybrid qubit system reads 
%
%
\begin{equation}
\widehat{H}_{2q}=\epsilon _{1}{(V_{g})}\widehat{\sigma }_{z}^{(1)}-\Delta
_{1}\widehat{\sigma }_{x}^{(1)}+\epsilon _{2}{(\phi _{2})}\widehat{\sigma }
_{z}^{(2)}-\Delta _{2}\widehat{\sigma }_{x}^{(2)}+\chi {\widehat{\sigma }
_{x}^{(1)}}{\widehat{\sigma }_{z}^{(2)}},  \label{Hhbr}
\end{equation}
where $\epsilon _{1}$ and $\Delta _{1}$ are the longitudinal and transversal
parameters of the charge qubit, and likewise $\epsilon _{2}$ and $\Delta _{2}$ for the flux qubit. The interaction between
the qubits gives rise to an effective $\sigma _{x}\otimes {\sigma _{z}}$ geometric term \cite{xiao} with strength $\chi$.

\section{Coupling to the bath}
In order to study the system's dissipative dynamics, we modeled the
dissipative environment as either one or two baths of harmonic oscillators
which are coupled to the $\sigma _{z}$ components of each qubit. The first
case considers each qubit coupled to its own bath of oscillators
in the form $\widehat{\sigma }_{z}^{(i)}\widehat{X}^{i}$, where $
\widehat{X}^{i}$ ($i=1,2$) denotes the bath coordinates. The full interacting Hamiltonian reads
\begin{equation}
\widehat{H}_{2B}=\sum_{i=1,2}\left( -\frac{1}{2}\epsilon _{i}{\widehat{
\sigma }_{z}^{(i)}}-\frac{1}{2}\Delta _{i}{\widehat{\sigma }_{x}^{(i)}}+
\frac{1}{2}{\widehat{\sigma }_{z}^{(i)}}\widehat{X}^{i}\right) +\chi {\sigma
_{x}^{(1)}}{\sigma _{z}^{(2)}}+\widehat{H}_{B1}+\widehat{H}_{B2},
\label{h2b}
\end{equation}
where $\widehat{H}_{Bi}$ ($i=1,2$) is the Hamiltonian of the $i$-th bath.
In the second case, we consider that  the qubits are coupled to a common bath. Hence,  the full system's
Hamiltonian reads
\begin{equation}
\widehat{H}_{1B}=\sum_{i=1,2}\left( -\frac{1}{2}\epsilon _{i}{\widehat{
\sigma }_{z}^{(i)}}-\frac{1}{2}\Delta _{i}{\widehat{\sigma }_{x}^{(i)}}
\right) +\frac{1}{2}\left( \widehat{\sigma }_{z}^{(1)}+\widehat{\sigma }
_{z}^{(2)}\right) \widehat{X}+\chi {\widehat{\sigma }_{x}^{(1)}}{\widehat{
\sigma }_{z}^{(2)}}+\widehat{H}_{B},  \label{h1b}
\end{equation}
where $\widehat{X}$ is the bath coordinate and $\widehat{H}_{B}$ represents
the Hamiltonian of the common bath. \\ \\
{\it Bloch-Redfield formalism.---} 
This formalism  is derived from a projector operator approach and
provides an important tool for finding a set of
coupled master equations which describe the dynamics of the reduced density
matrix for a given system in contact with a dissipative environment. The Liouville equation for the total density operator $\rho _{T}$ (of the whole
system) is given by  $
i\hbar \frac{d}{dt}\rho _{T}(t)=[ \widehat{H}(t),\rho _{T}(t)]$.
%
 The corresponding reduced density
matrix obeys, in the Born-Markov limit, the so-called Redfield equation 
\cite{weis}: 
\begin{equation}
\dot{\rho}=-i\omega _{nm}\rho _{nm}(t)-\sum_{kl}R_{nmkl}\rho _{kl}(t),
\label{den}
\end{equation}
where the first term represents the reversible motion in terms of the transition
energies $ \hbar\omega _{nm}\equiv  E_{n}-E_{m}$, and the
second term describes the system's relaxation. The
Redfield relaxation tensor $R_{nmkl}$   includes the dissipative efects of the system-environment coupling:
\begin{equation}
R_{nmkl}=\delta _{lm}\sum_{r}\Gamma _{nrrk}^{(+)}+\delta _{nk}\sum_{r}\Gamma
_{lrrm}^{(-)}-\Gamma _{lmnk}^{(+)}-\Gamma _{lmnk}^{(-)},  \label{tensor}
\end{equation}
and its  components are given by the `golden rule'
expressions 
\begin{eqnarray}
\Gamma _{lmnk}^{(+)} &=&\hbar ^{-2}\int_{0}^{\infty }{dte^{-i\omega
_{nk}t}\left\langle {\tilde{H}_{I,lm}(t)\tilde{H}_{I,nk}(0)}\right\rangle },
\nonumber \\
\Gamma _{lmnk}^{(-)} &=&\hbar ^{-2}\int_{0}^{\infty }{dte^{-i\omega
_{lm}t}\left\langle {\tilde{H}_{I,lm}(0)\tilde{H}_{I,nk}(t)}\right\rangle }.
\label{rta}
\end{eqnarray}
Here, $\tilde{H}_{I}(t)=\exp ( \frac{i\widehat{H}_{B}t}{\hbar }) {
H_{I}}\,{\exp( \frac{-i\widehat{H}_{B}t}{\hbar }) }$ is the
interaction term in the interaction picture, and the bracket denotes the thermal
average over the bath's degrees of freedom.


\section{Results and Discussion}


{\it System's eigen-energies.---}
These are worked out in the singlet/triplet
basis states. In this basis, the Hamiltonian $H_{2q}$ of the two-qubit system  assumes the matrix form:
\begin{eqnarray}
\widehat{H}_{2q} = -\frac{1}{2}\left(\begin{array}{c c c c } \epsilon & \eta + \chi &0&-\chi  \\ \eta + \chi & 0 &\eta - \chi &0  \\ 0&\eta - \chi &  - \epsilon& -\chi \\ -\chi& 0& -\chi& 0    \end{array}\right) \ ,
\end{eqnarray}
where we have introduced $\epsilon = \epsilon_{1} + \epsilon_{2}$, $\Delta{\epsilon}= \epsilon_{1} - \epsilon_{2}$, $\eta = \frac{\Delta_{1} + \Delta_{2}}{\sqrt{2}}$, $\Delta{\eta} = \frac{\Delta_{1} - \Delta_{2}}{\sqrt{2}}$, and considered qubits with the same transition energies $\epsilon_{1}=\epsilon_{2}$ and transversal coupling $\Delta_{1}=\Delta_{2}$. Thus, $\Delta{\epsilon}=\Delta{\eta}=0$.
\begin{figure}[h]
\begin{center}
\includegraphics[width=1\textwidth]{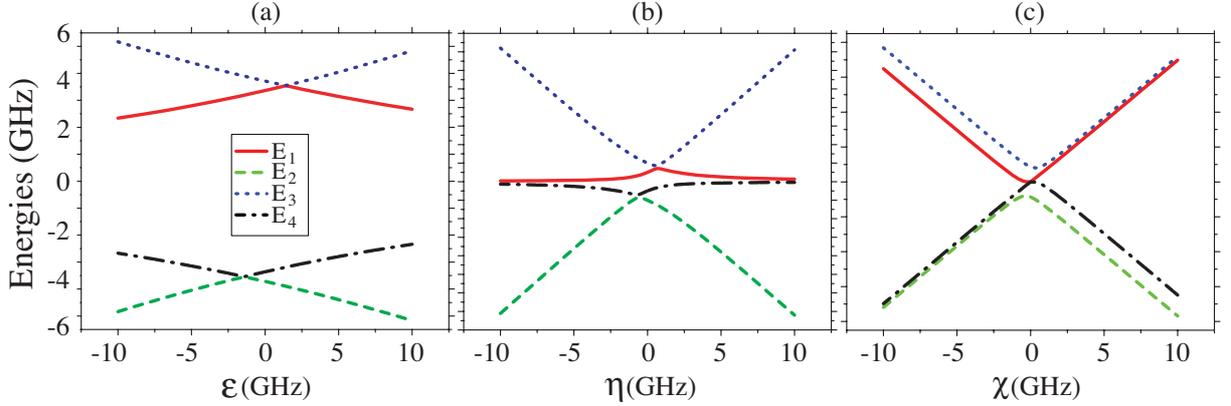}
\end{center}
\vspace{-0.5cm}
\caption{Energy spectrum of the coupled hybrid two-qubit system.  (a)  $\chi=10$ GHz, $\eta=1$ GHz, and $\epsilon $ is varied, (b) $\chi=1$ GHz, $\epsilon=1$ GHz ($\eta$ is varied), and (c) $\epsilon=\eta=1$ GHz  ($\chi$ is varied). In all the graphs, $E_{1}$ is the singlet state and $E_{2},E_{3}$, and  $E_{4}$ are the triplet states.}
\label{fig1}
\end{figure}

In Figure \ref{fig1}, we present the system's energies  for three
different scenarios, in which two of the parameters $\epsilon$, $\eta $, $\chi $ have been fixed  and the third one is varied. In Figure \ref{fig1}(a) we
have varied the longitudinal parameter $\epsilon $, showing  a crossing
between the energy of the singlet ($E_{1}$) and one of the energies of the
triplet ($E_{3}$). In Figure \ref{fig1}(b) the influence of the transversal
parameter $\eta $ has been plotted. We notice the appearance of the same crossing observed in Figure \ref{fig1}(a), between the singlet and the $E_{3}$--triplet
state, which indicates that these states do not interact. 
Figure \ref{fig1}(c) shows the dependence on the coupling term $\chi$. Now, a crossing between the energy of the singlet state and that of the triplet state $E_{4}$ takes place. Note that for large values of $\chi $, there is a degeneracy between the singlet
state $E_{1}$ and the triplet state $E_{3}$. This feature means that for large
positive values of the parameter $\chi$, a large ferromagnetic
coupling is present between the qubits. 
\\

\noindent
{\it  Dissipative dynamics and temperature effects.---}
The solution of the Redfield equation $\dot{\rho}_{nm}=-\sum_{kl}R_{nmkl}\rho _{kl}(t)$
allows to analyse the system's reduced dynamics, for which explicit
expressions have been found for the components of Equation (\ref{rta}), for both types of
coupling to the environment.
 In the case of  independent coupling we get 
\begin{equation}
\Gamma _{lmnk}^{(+)}=\Gamma _{lmnk}^{(-)}=\frac{1}{4\beta {\hbar }}\left[ {
\sigma _{z,lm}^{(1)}\sigma _{z,nk}^{(1)}\alpha _{1}+\sigma
_{z,lm}^{(2)}\sigma _{z,nk}^{(2)}\alpha _{2}}\right] ,  \label{r2b}
\end{equation}
where the $\alpha_i$'s correspond to the bath coupling parameters for the specific cases of Ohmic spectral densities $J_{i}(\omega )=\alpha _{i}\hbar \omega /\Big(1+\frac{\omega ^{2}}{\omega
_{c}^{2}}\Big)$, and
$\sigma _{z,ij}\equiv \left\langle E_{i}\right\vert \sigma _{z}\left\vert
E_{j}\right\rangle$. 
  For the case of collective coupling, $J(\omega )=\alpha \hbar \omega /\Big(1+\frac{\omega ^{2}}{\omega
_{c}^{2}}\Big)$, and the rates become
\begin{equation}
\Gamma _{lmnk}^{(+)}=\Gamma _{lmnk}^{(-)}=\frac{\alpha }{4\beta {\hbar }}
\left[ {\sigma _{z,lm}^{(1)}\sigma _{z,nk}^{(1)}+\sigma _{z,lm}^{(1)}\sigma
_{z,nk}^{(2)}+\sigma _{z,lm}^{(2)}\sigma _{z,nk}^{(1)}+\sigma
_{z,lm}^{(2)}\sigma _{z,nk}^{(2)}}\right] .  \label{r1b}
\end{equation}
\begin{figure}[h]
\begin{center}
\includegraphics[width=1.0\textwidth]{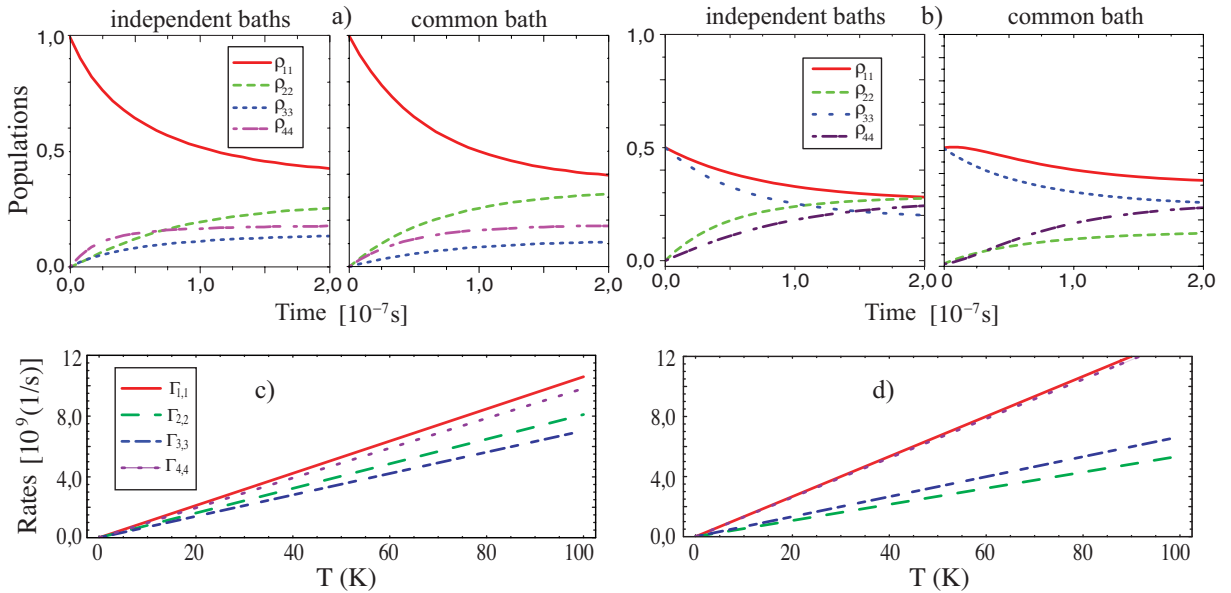}
\end{center}
\vspace{-0.5cm}
\caption{Populations dynamics, $\protect\rho _{ii}(t)$. The  density matrix calculation has been performed for the initial conditions: a) pure
state $|E_{1}\rangle $, and b) maximally entangled Bell state 
$|\Phi ^{+}\rangle $. Lower graphs show the relaxation rates (see main text)
as a function of the bath temperature for the cases of c) independent 
and d) collective coupling to the environment. $\alpha_i=\alpha=10^{-3}$.}
\label{fig2}
\end{figure}

The reduced density matrix has been computed from  the system of
differential equations  (\ref{den}). Figure \ref{fig2} shows the time dependence of the system's populations $\rho_{ii}$ for a  coupling to two different kinds of 
environments: i) independent and ii) collective or common baths. Figures 3(a) and 3(b) correspond to the  initial conditions $\left| E_{1}\right\rangle$, and  $\left|\Phi^{+}\right\rangle = \frac{1}{\sqrt{2}}\left(\left|00\right\rangle + \left|11\right\rangle\right)$, respectively. The population $\rho _{11}$ has a slightly faster decay in the case of 
collective coupling for the initial condition  $\left| E_{1}\right\rangle$, whereas this behaviour is reversed for the case of the entangled initial condition $\left|\Phi^{+}\right\rangle$. Thus, the dissipative population dynamics is very sensitive to initial state preparations.

\begin{figure}
\begin{center}
\includegraphics[width=1\textwidth]{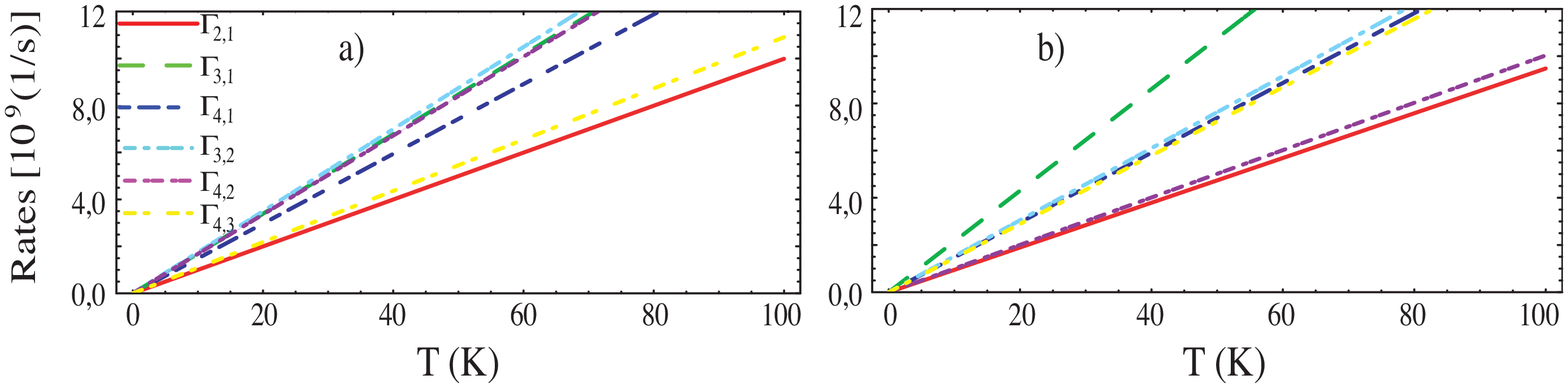}
\end{center}
\vspace{-0.5cm}
\caption{Decoherence rates as a function of the reservoir temperature for: a) independent and b) collective bath coupling.  $\alpha_i=\alpha=10^{-3}$.}
\label{fig3}
\end{figure}

The behaviour of the relaxation rates 
$\Gamma _{{n,n}}^{i/c}=Re({R_{nnnn}^{i/c}})$ 
as a function of the bath temperature
is shown in Figures \ref{fig2}(c) and (d), for the initial conditions $\left| E_{i}\right\rangle$. The superscripts 
$i$, and $c$ stand for  independent (Fig. \ref{fig2}(c)) and collective (Fig. \ref{fig2}(d)) couplings respectively. 
In agreement with graph 3(a), this shows that
the rate $\Gamma _{11}$ tends to a higher value for the case of collective
coupling.
In Figure \ref{fig3}, we have plotted the dephasing (decoherence)
rates 
$
\Gamma _{{n,m}}^{i/c}=Re({R_{nmnm}^{i/c}})$, as a function of the bath temperature, for the cases of a) independent
and b) collective coupling. We
note that, in both cases, the decoherence rates are of the same order of magnitude. Furthermore, it
is observed that $\Gamma _{21}$ is the minimum dephasing rate, while $\Gamma _{31}$ displays the higher rate.

By comparing the decoherence and the relaxation rates obtained, we conclude that both types of dissipation mechanisms occur on the same time scale. This implies a long decoherence time, which is a key requirement  in the construction of quantum devices that are expected to work as quantum information processors.
\vspace{-0.2cm}
\ack This work was partially supported by
Colciencias grant 1106-452-21296, the
Excellence Centre for Novel Materials (CENM), and the  exchange program PROCOL (DAAD-Colciencias).

\end{document}